# Relaxation dynamics in the one-dimensional organic charge-transfer salt $\delta$-(EDT-TTF-CONMe$_2$)$_2$Br

J. K. H. Fischer[1,*,#], P. Lunkenheimer[1], C. Leva[2], S. M. Winter[2,3], M. Lang[2], C. Mézière[4], P. Batail[4], A. Loidl[1] and R. S. Manna[2,5,#]

[1]*Experimental Physics V, Center for Electronic Correlations and Magnetism, University of Augsburg, 86159 Augsburg, Germany*
[2]*Institute of Physics, Goethe University Frankfurt (M), Max-von-Laue-Str. 1, 60438 Frankfurt (M), Germany*
[3]*Institute for Theoretical Physics, Goethe University Frankfurt (M), Max-von-Laue-Str. 1, 60438 Frankfurt (M), Germany*
[4]*MOLTECH-Anjou, UMR 6200, CNRS-Université d'Angers, Bâtiment K, Angers F-49045, France*
[5]*Department of Physics, IIT Tirupati, Tirupati 517506, India*

A detailed investigation of the charge-ordered charge-transfer salt $\delta$-(EDT-TTF-CONMe$_2$)$_2$Br by thermal-expansion measurements and dielectric spectroscopy reveals three dynamic processes of relaxational character. The slowest one exhibits the characteristics of glassy freezing and is ascribed to the conformational dynamics of terminal ethylene groups of the organic molecules. Such a process was previously found for related charge-transfer salts where, however, the anions form polymerlike chains, in contrast to the spherical anions of the present material. Dielectric spectroscopy reveals two additional relaxational processes. The characteristics of the faster one are consistent with excitations of a one-dimensional Wigner lattice as recently observed in this material by infrared spectroscopy, which are also accompanied by conformational changes of the molecules. However, at low temperature the ethylene-group relaxation exhibits the cooperativity-induced dramatic slowing down that is typical for glassy freezing, while the defect-related Wigner-lattice excitation follows thermally activated behavior as expected for single-dipole relaxations.

## I. INTRODUCTION

Among the organic charge-transfer salts, there are some interesting examples of strongly correlated electron systems in reduced dimensions [1]. These materials are attracting further attention as some of them exhibit ferroelectricity that most likely is of electronic origin [2,3]. Here the polar order arises from electronic degrees of freedom, instead of the more common off-center displacement of ions found in conventional ferroelectric materials. Electronic ferroelectricity has recently come into the focus of interest as it is a promising mechanism for the development of new ferroelectric devices and a prominent way to generate multiferroicity [2,4,5]. The key phenomenon for electronic ferroelectricity is charge order (CO), which is controlled by strong Coulomb correlations. Clear examples of CO have been found in a variety of organic charge-transfer salts with effectively ¼-filled hole bands, which thus are good candidates for electronic ferroelectricity [2,3,6,7,8,9,10,11,12,13,14] and even multiferroicity [11,14,15]. One of the outstanding recent examples is the dimerized salt $\kappa$-(BEDT-TTF)$_2$Cu[N(CN)$_2$]Cl (hereafter abbreviated $\kappa$-Cl), where BEDT-TTF stands for bis(ethylenedithio)-tetrathiafulvalene (often abbreviated as ET). It was suggested to exhibit multiferroicity due to the simultaneous occurrence of CO-driven ferroelectricity and magnetic order [11]. However, while the existence of ferroelectric and antiferromagnetic order in $\kappa$-Cl has been unambiguously demonstrated, the presence of CO in this material is still controversially discussed [16,17,18]. In contrast, for the related compound $\kappa$-(BEDT-TTF)$_2$Hg(SCN)$_2$]Cl ($\kappa$-Hg) charge-order has been clearly identified by vibrational spectroscopy [19]. Moreover, indications for electronically-driven ferroelectricity have been found recently [20]. Another interesting recent example is $\alpha$-(ET)$_2$I$_3$, which shows the signature of relaxor-ferroelectricity [21]. In this compound the metal-insulator transition below $T_{CO} \approx 135$ K is a well-established fact [22,23,24,25].

In the present work, we investigate the organic charge-transfer salt $\delta$-(EDT-TTF-CONMe$_2$)$_2$Br ($\delta$-Br), where EDT-TTF-CONMe$_2$ [4,5-ethylenedithio-4'-(N,N-dimethyl-carbamoyl)tetrathiafulvalene] represents the donor-molecule and Br is the counter anion. Since this compound exhibits CO [26,27,28], which exists already at room temperature, and as its space group was reported to be compatible with ferroelectricity [26], at first glance it may be considered a candidate for showing electronic ferroelectricity. However, $\delta$-Br lacks dimerization [26] and, thus, the special ferroelectricity mechanism related to intra-dimer degrees of freedom that was considered for the mentioned compounds $\kappa$-Cl, $\kappa$-Hg, and $\alpha$-(ET)$_2$I$_3$ cannot play a decisive role in this material. In the present work, we report a thorough dielectric investigation of $\delta$-Br checking for the expected differences in dielectric behavior compared to the former compounds and looking for possible dipolar dynamics in this charge-ordered material. Furthermore, using dielectric and thermal-expansion measurements, we search for possible signatures of the glasslike relaxation process that was found in several related charge-transfer salts [29,30,31,32]. It is thought to mirror the glassy freezing of the configurational dynamics of the organic molecules but other contributions were also discussed [33,34]. The associated structural disorder is believed to play an important role for the electronic ground-state properties of these systems.

Previous work on $\delta$-Br has shown that it exhibits a charge-ordered Mott insulating state at room temperature and ambient pressure [26,27,28]. Upon cooling, at $T_s = 190$ K a structural orthorhombic-to-monoclinic phase transition occurs. At about 12 K, antiferromagnetic order was reported [27,35]. It was speculated that there may be an additional weak ferromagnetic contribution at low temperatures caused by Dzyaloshinskii-Moriya exchange interaction [36]. By means of infrared measurements [37], an optical gap of 68 meV was estimated and ascribed to domain-wall excitations as expected for one-dimensional Wigner lattices [38,39,40]. Moreover, in Ref. [37] $\delta$-Br was found to remain quasi-one-dimensional from room temperature down to 10 K. The formation of the conduction band in this material is due to a $\pi$-$\pi$ overlap between the donor molecules. These molecules are stacked equidistantly along the crystallographic $a$-axis forming a charge-ordered chain of nearly neutral and positively charged molecules [26]. These stacks are adjacent along the $b$ axis and separated by Br anions in the $c$ direction, leading to the formation of planes of EDT-TTF-CONMe$_2$ molecules perpendicular to the $a$ axis. Within these planes, a CO pattern was also found along the $b$ direction [26,28]. It should be noted, however, that the interstack distance is significantly larger than the intermolecular distance within a single stack making the material quasi-one-dimensional. A rather high charge disproportionation ratio for two neighboring molecules of about 9:1 was detected [26,27,28]. The lack of dimerization makes $\delta$-Br a prototypical example of a one-dimensional Mott insulator with quarterfilled conduction band [26,27,28,37,41].

In the present work, by measurements of the uniaxial thermal-expansion coefficients of $\delta$-Br over a wide range of temperature, we find anomalous lattice effects reflecting the second-order structural phase transition from orthorhombic to monoclinic at 190 K. Another anomaly is detected in the temperature range 110 K to 130 K, whose dependence on the cooling rate reveals its relaxational nature. We ascribe it to motions of the terminal groups in the asymmetric EDT-TTF-CONMe$_2$ molecules. Furthermore, by dielectric spectroscopy over a wide temperature and frequency range we detect two additional relaxational processes. A likely relation to domain-wall excitations within the Wigner lattice found for $\delta$-Br by infrared spectroscopy [37] is discussed.

## II. EXPERIMENTAL DETAILS

Crystals of $\delta$-Br were grown as reported in Ref. [26]. The geometry of the investigated samples was needle-like ($a$-axis parallel to the long direction).

The sample examined by dielectric spectroscopy was a needle-like single crystal, with a size of 3.6 mm along the $a$-axis and ~0.6 mm along the $b$- and $c$-axes. For the dielectric measurements, contacts of graphite paste formed like caps were applied on either side of the crystal, ensuring an electric-field direction parallel to the $a$-axis. The dielectric constant and conductivity were determined using a frequency-response analyzer (Novocontrol Alpha-A). Sample cooling was achieved by a $^4$He-bath cryostat (Cryovac).

Thermal-expansion measurements were performed by using an ultrahigh-resolution capacitive dilatometer, built after Ref. [42], enabling the detection of length changes $\Delta l \geq 10^{-2}$ Å, where $l$ is the length of the sample. The measurements were performed on a single crystal of $\delta$-Br along the crystallographic $a$ axis. The size of the sample along the $a$ and $b$ axes were 1.225 mm and 0.16 mm, respectively. The crystal was measured after confirming the orientation of the crystallographic axes of the bulk sample by Laue diffraction. To ensure thermal equilibrium the measurements were performed upon heating with slow sweep rates of 1.5 K/h ($T < 30$ K) and 3 K/h ($T > 30$ K), after precooling with different rates.

## III. RESULTS AND DISCUSSION

### A. Dielectric spectroscopy

Figure 1 shows the temperature-dependence of the dielectric constant $\varepsilon'(T)$ and conductivity $\sigma'(T)$ measured at various frequencies. For better visibility, only a selection of frequencies is shown for $\varepsilon'(T)$. In the conductivity [Fig. 1(b)], two peaks are revealed whose peak temperatures strongly depend on frequency as indicated by the dashed lines. They are nearly merged at low and well separated at high frequencies. The lower-temperature peaks are accompanied by clear steps in the permittivity [Fig. 1(a)] with an amplitude $\Delta\varepsilon$ of the order of 30. A closer inspection of $\varepsilon'(T)$ in the region around 100 - 200 K reveals steps ($\Delta\varepsilon \approx 5$) corresponding to the high-temperature peaks, too (inset of Fig. 1). They are superimposed to a general rise of $\varepsilon'$ with increasing temperature. Generally, steps in the real part of the dielectric constant and accompanying peaks in the conductivity (which is directly related to the dielectric loss via $\varepsilon'' \propto \sigma'/\nu$), suggest relaxational behavior typically arising from the reorientational motion of dipolar degrees of freedom [43,44,45]. For a given measurement frequency $\nu$, at the peak temperatures $T_p$ the temperature-dependent relaxation time $\tau(T)$ matches the condition $\tau(T_p) = 1/(2\pi\nu)$. The observed shift of the spectral features to lower temperatures with decreasing frequency mirrors the slowing down of the relaxation time with decreasing temperature. Very similar behavior is commonly found for the glassy freezing of molecular motion in dipolar glass formers [43,44,45]. The two relaxation processes revealed in Fig. 1 will be termed process I (high-temperature relaxation) and II (low temperature) in the further course of this work.

In Fig. 1(b), at the lower frequencies the high-temperature peak associated with process I is barely visible and immediately succeeds the low-temperature one (e.g., in case of the 1 Hz curve at about 48 K and $1\times10^{-12}$ $\Omega^{-1}$cm$^{-1}$). It becomes rather well separated starting with the 184 Hz curve and gets very broad in the kHz range before finally shifting out of the experimentally accessed window at the highest frequencies. Interestingly, between 33.8 kHz and 140 kHz an unusual crossing of the curves of process I is observed around 150 K. When examining the development of the $\sigma'(T)$ curves with decreasing frequency, this crossing seems to be mainly due to an anomalous behavior of the local minimum, which strongly broadens between 140 and 54.4 kHz. This could indicate an additional contribution from a third relaxation process, which, however, is not clearly resolved.

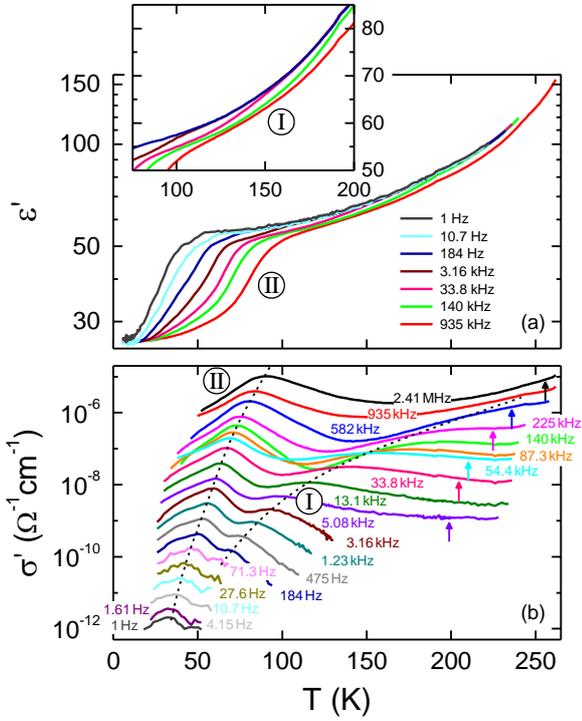

FIG. 1. Temperature dependence of the dielectric constant $\varepsilon'(T)$ (a) and conductivity $\sigma'(T)$ (b) of $\delta$-Br as measured for various frequencies along the $a$ axis. The two observed relaxation processes are marked by I and II; the development of the corresponding loss-peak positions is indicated by the dashed lines in (b). For clarity, $\varepsilon'(T)$ is only presented for seven frequencies instead of the 20 shown in $\sigma'(T)$. The conductivity curves were cut at low and high temperatures where data scatter and systematic errors start to dominate the data. For selected frequencies, the arrows in (b) indicate the expected peak positions arising from the glasslike relaxation as calculated from the VFT curve shown in Fig. 4. The inset shows a zoomed view of $\varepsilon'(T)$ revealing the superimposed step associated with the high-temperature relaxation I.

Finally, the structural transition at 190 K apparently does not influence the dielectric properties of $\delta$-Br. While the conductivity becomes too small below about 25 K to be properly measured [Fig. 1(b)], $\varepsilon'(T)$ could be detected down to 5.6 K but reveals no signature of the magnetic transition at 12 K [Fig. 1(a)].

The mentioned general rise of $\varepsilon'$ with increasing temperature by nearly a factor of three, observed at temperatures above about 100 K in Fig. 1(a), is unusual and currently we do not have an explanation for this behavior. Such non-canonical temperature dependence of $\varepsilon'$ can be found below a polar phase transition, reflecting the growth of polar domains with decreasing temperature. Thus one may speculate about such a transition in $\delta$-Br at elevated temperatures. Indeed, the CO was reported to arise at about 700 K in this material [27]. Dielectric measurements at elevated temperatures are necessary to clarify this issue.

## B. Thermal Expansion

Figure 2 shows the uniaxial expansion coefficients $\alpha_a = l_a^{-1}\,dl_a/dT$ measured along the crystallographic $a$-axis below 200 K for $\delta$-Br. The overall expansivity along the $a$-axis is very large. $\alpha_a(T)$ shows a pronounced anomaly at around 190 K, the shape of which confirms a second-order phase transition, consistent with Ref. [26]. It is related to the structural phase transition from the orthorhombic to the monoclinic structure and is accompanied by twinning [26]. Towards lower temperatures the expansivity decreases almost linearly followed by another, steplike anomaly around 120 K, which we ascribe to a glasslike transition.

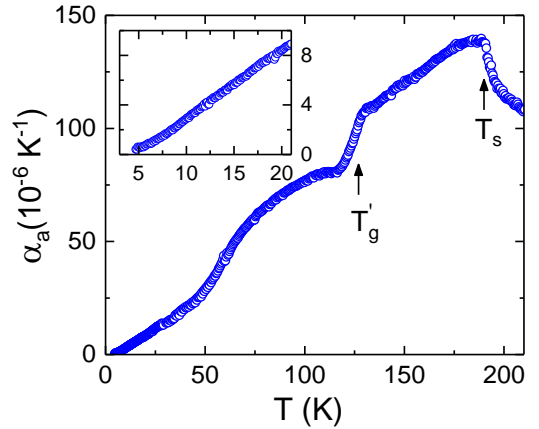

FIG. 2. Temperature dependence of the uniaxial thermal expansion coefficient $\alpha_a$ of $\delta$-Br measured along the $a$-axis under heating after cooling with 10 K/h. The structural and glasslike transitions are indicated by the arrows. The inset shows the data across the antiferromagnetic transition, expected at 12 K, on an expanded scale.

To check for the glasslike nature of this anomaly, measurements with different thermal history were performed.

In contrast to conventional phase transitions, glass transitions generally should depend on the thermal history of the sample. Especially, it is well established that the anomalies observed in various quantities under cooling at the glass transition (e.g., specific heat, thermal expansion, dielectric constant) shift to higher temperatures for higher cooling rates [46]. Thus, one may regard the actual glass temperature $T_g'$ as dependent on the cooling rate. (To avoid ambiguities, $T_g$ as a material property is usually defined for a fixed cooling rate of 10 K/min. Therefore, here we denote the rate dependent glass temperature as $T_g'$.) However, to ensure thermal equilibrium the present thermal expansion measurements could not be performed under cooling with significantly different cooling rates $q_c$. Instead, the sample was precooled with three different rates between -3 and -23 K/h and measured afterwards under heating with a fixed rate of +3 K/h. It is well known that, in case of a glass transition, this should also lead to a temperature shift of the anomaly because, in principle, different types of glasses are prepared by different precooling rates The resulting $\alpha_a(T)$ curves in the region of the transition are shown in Fig. 3. Indeed, the anomaly is shifted by several kelvins for the different precooling rates [46]. It occurs at the highest temperature for the fastest rate, just as expected for a glass transition. A slight undershoot of $\alpha_a$ is observed for the fastest precooling rate as indeed expected for heating rates that are much slower than the preceding cooling [46].

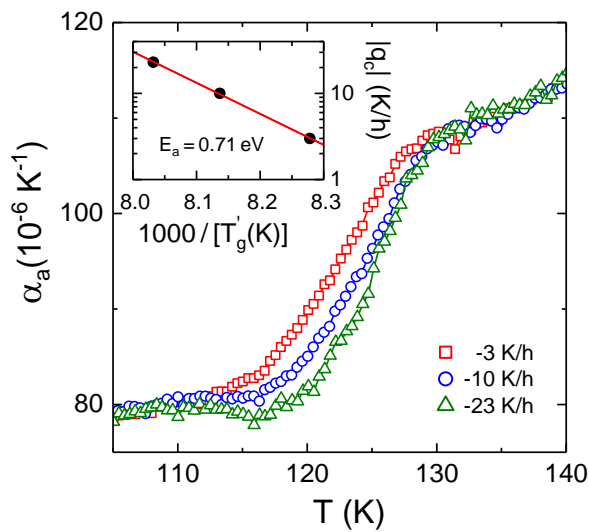

FIG. 3. Thermal expansion coefficient along the $a$-axis across the glasslike transition on an expanded scale, measured with a heating rate of +3.0 K/h after cooling with three different rates. The inset shows an Arrhenius plot of the cooling rate $|q_c|$ vs the temperature of the glasslike anomaly determined by reading off the midpoint of the broad rise in $\alpha_a(T)$. The line is a linear fit, corresponding to thermally activated behavior.

Similar features in thermal-expansion data have previously been observed in the quasi-two-dimensional organic charge transfer salts $\kappa$-(BEDT-TTF)$_2$X with X = Cu[N(CN)$_2$]Cl ($\kappa$-Cl), Cu[N(CN)$_2$]Br ($\kappa$-Br), Cu(NCS)$_2$ ($\kappa$-CuNCS) [31], and Hg(SCN)$_2$Cl ($\kappa$-Hg) [47]. They were attributed to the freezing of conformational degrees of freedom of the terminal ethylene groups of the ET molecules, which obviously influence the volume of the crystal to be detectable by thermal expansion [31]. Via short C-H···anion contacts, the anions were assumed to be involved to some extent in the freezing of the ethylene end groups, which clearly is not a purely intramolecular effect [31,34]. In the present case of $\delta$-Br, to our knowledge such a glasslike transition was observed for the first time in a quasi-one-dimensional system with a discrete, spherical anion, in contrast to the aforementioned quasi-two-dimensional compounds, where the anions form polymerlike chains.

It should be noted that the origin of the glasslike transition in the mentioned systems [31] is still controversial. Recent synchrotron x-ray diffraction experiments [33,48] established the level of residual ethylene group disorder in $\kappa$-Br. From this the authors concluded that the glasslike transition is not primarily caused by the configurational freezing-out of the ET-endgroup motions, but rather other structural and electronic degrees of freedom have to be taken into account. Our results on $\delta$-Br with its simple spherical anion might help to clarify the origin of this glasslike feature. For example, it was speculated that the glasslike transition may involve reorientational motions of the polymeric anion chains [33], which can be clearly excluded for $\delta$-Br.

The performed thermal-expansion experiments, where the different precooling rates provide the time scale of the experiment, in principle reveal information on the relaxational dynamics of the glass-forming system. These data can be analyzed via an Arrhenius plot of the cooling rate $|q_c|$ (which can be assumed to be proportional to the inverse relaxation time) vs the inverse temperature of the anomaly [49,50] (inset of Fig. 3). Linear behavior shows up indicating thermally-activated slowing down of the relaxation dynamics under cooling. From a linear fit (line), the activation energy $E_a$, corresponding to the slope in the Arrhenius plot, can be determined [31,50]. It yields an activation energy of 0.71 eV. This is significantly larger than the values $E_a$ = 0.28 eV, 0.23 eV, and 0.23 eV obtained for $\kappa$-Br, $\kappa$-Cl, and $\kappa$-Hg respectively [31,47].

Since the asymmetric EDT-TTF-CONMe$_2$ molecules in $\delta$-Br also contain ethylene end groups, similar to the systems treated in Ref. [31], we speculate that the detected glasslike transition in the present material is caused by the freezing of their conformational motions as well. In $\delta$-Br, the activation barrier for such conformational motion may be affected by strong electrostatic and Van-der-Waals coupling between the ethylene groups and the heavy Br anions as well as coupling between the ethylene groups and the –CONMe$_2$ moiety on adjacent molecules, which are dominated by short (< 2.5 Å) CH···O contacts. In order to estimate the effect of these interactions on $E_a$, we performed ab-initio quantum chemistry calculations as discussed in Ref. [34], using the ORCA program [51] at the BP86 and B3LYP/def2-SV(P) levels.

Calculations were performed on single molecules with starting geometries based on the monoclinic structure of Ref. [26]. Coupling of the ethylene groups to the adjacent –CONMe$_2$ and the Br anions was included via static OPLS-aa forcefields [52]. The ethylene groups were relaxed in order to estimate the relative energies of different conformations. In all cases, the charge of the donor molecules was found to have little effect. It should be noted that this approach does not explicitly consider the role of cooperativity between different ethylene groups, which should enhance the effective activation barriers near the glass transition. As a result, experimental $E_a$ values obtained from the cooling-rate dependence of $T_g$' (as above) are systematically found to be larger by a factor of ~1.3 to 1.5 when compared to the ab-initio estimates [34,47].

Using this approach, we find that the –CONMe$_2$ moieties play the dominant role in confining the motion of the ethylene groups, leading to an estimated $E_a$ on the order of ~0.5 eV, which is consistent with the experimental value of 0.71 eV. The computed $E_a$ values are surprisingly insensitive to the presence or absence of the Br, suggesting that the anions play only a minor role in the glass transition. In contrast, omission of the coupling to the –CONMe$_2$ groups leads to a strong reduction of $E_a$ by nearly half. This observation likely explains the much larger $T_g$' and $E_a$ measured in $\delta$-Br, compared to the $\kappa$-phase BEDT-TTF salts. Finally, it is interesting to note that the energy differences $2\Delta E$ between the two metastable ethylene conformations were computed to be $\Delta E/E_a \sim 10$ in $\delta$-Br, which satisfies the empirical rule of $\Delta E/E_a > 5$ suggested in Ref. 34 for the existence of a glass transition in $\kappa$-phase BEDT-TTF salts.

Cooling below the suggested glasslike transition, the thermal-expansion data of Fig. 2 show a smeared-out anomalous lattice effect of unknown origin in the temperature range of 25 K to 60 K. Finally, just as in the dielectric measurements, the expansion data do not exhibit any signature of the antiferromagnetic transition around 12 K (see inset of Fig. 2) reported in Refs. [27,35].

### C. Relaxation times

Figure 4 shows an Arrhenius representation of the temperature-dependent relaxation times $\tau$ of the two processes revealed by dielectric spectroscopy (squares and triangles) and of the process detected by thermal expansion (circles). The dielectric-spectroscopy data indicated by empty symbols were deduced from the loss-peak positions in the temperature-dependent graph [Fig. 1(b)], via the relation $\tau(T_p) = 1/(2\pi\nu)$, while the filled symbols were determined from fits to spectra of $\varepsilon''$ for different temperatures (not shown). The found slight deviations of the relaxation times in these two cases arises from small shifts in the peak positions of the temperature-dependent data, which are caused by differences in peak width and height at different temperatures that distort the temperature-dependent curves.

In order to calculate relaxation times from the thermal-expansion data, we assumed the common definition of the glass transition temperature $T_g$ as the temperature, at which the system falls out of equilibrium for a precooling rate of $q_c = -10$ K/min. When considering that the glass transition typically occurs at a relaxation time of about 250 s [53], we can assign the three precooling rates (-3 K/h, -10 K/h, and -23 K/h) to corresponding relaxation times $\tau \propto 1/|q_c|$ (50000 s, 15000 s, and 6522 s, respectively). Previously, similar methods have been used to deduce relaxation times from cooling rates [49,50,54,55].

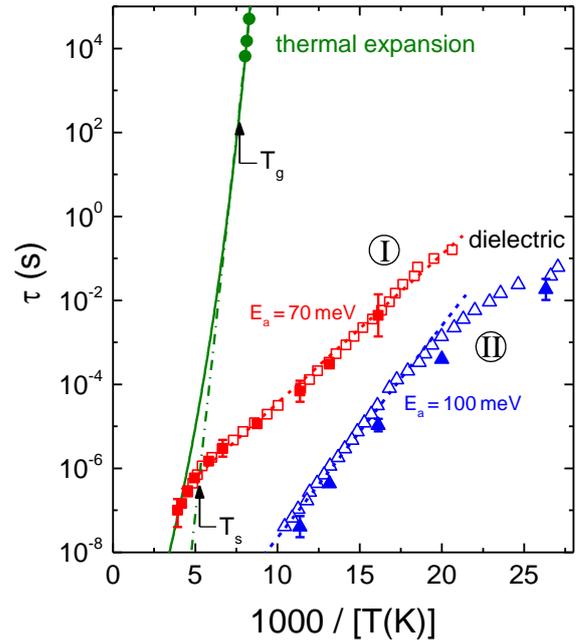

FIG. 4. Arrhenius plot of the temperature-dependent relaxation times determined by dielectric spectroscopy (squares: process I, triangles: process II) and thermal expansion (circles). For the dielectric data, the filled symbols show the results obtained from fits to the frequency-dependent data; the empty symbols were deduced from the peak positions in the temperature-dependent graph [Fig. 1(b)]. The dashed lines indicate Arrhenius behavior with activation energies of 70 and 100 meV for the slower process I and faster process II, respectively. The dash-dotted line shows an Arrhenius fit ($E_a \approx 0.71$ eV) of the thermal-expansion data leading, however, to an unrealistic prefactor of $\tau_0 \approx 10^{-25}$ s. The solid line indicates VFT behavior with a reasonable attempt frequency of $3\times10^{-13}$ s. The structural and glasslike transitions are indicated by the arrows, and labeled $T_s$ and $T_g$, respectively.

When fitting the relaxation times as deduced from thermal expansion by Arrhenius behavior, $\tau = \tau_0 \exp[E_a/(k_BT)]$, implying thermally activated dynamics (dash-dotted line in Fig. 4), an activation energy of $E_a \approx 0.71$ eV and a prefactor of $\tau_0 \approx 10^{-25}$ s are obtained. As expected, the magnitude of $E_a$ agrees with that deduced from the inset of Fig. 3 because $|q_c| \propto 1/\tau$. However, as revealed by the evaluation within the relaxation-time representation, the obtained $\tau_0$ corresponds to an unreasonably high attempt frequency $\nu_0 = 1/(2\pi\tau_0)$ of about $10^{24}$ Hz. This indicates that deviations from Arrhenius behavior are likely to arise which, at lower $\tau$ values and higher temperatures, would lead to a slight bending of the log[$\tau(1/T)$] curve. However, at those lower $\tau$ values, experimental data do not exist due to the limitations of the rate-dependent expansion measurements. Indeed, such non-Arrhenius behavior is a well-known phenomenon in glass physics and commonly found for the main, structural relaxation process ($\alpha$ process) in many glass-forming liquids [44,45]. The solid line in Fig. 4 was calculated with the Vogel-Fulcher-Tammann (VFT) law [56], $\tau = \tau_0 \exp[B/(T-T_{VF})]$, with a reasonable $\tau_0$ value of $3\times10^{-13}$ s. This phenomenological formula is well established for the description of non-Arrhenius behavior in canonical glass formers [43,44]. Thus, the unrealistically high attempt frequency obtained from the Arrhenius fit strongly supports the relation of the detected relaxation process to glassy freezing. If applying the mentioned criterion $\tau(T_g) \approx 250$ s, based on the solid line in Fig. 4 we estimate a glass temperature of about 130 K. Using the Arrhenius fit (dash-dotted line) instead, leads to an only marginally higher $T_g$ of 131 K.

The relaxation times detected by dielectric spectroscopy (squares and triangles in Fig. 4) are many orders of magnitude faster than those deduced from the thermal-expansion measurements. For both dielectric processes, their temperature dependence can be reasonably well fitted by straight lines in a rather large $\tau$ range (dashed lines in Fig. 4) implying Arrhenius behavior. Process I has an activation energy $E_a$ of about 70 meV and a pre-exponential factor of $\tau_0 \approx 10^{-8}$ s. For the faster process II, $E_a \approx 100$ meV and $\tau_0 \approx 10^{-13}$ s are obtained.

Remarkably, the activation energy $E_a \approx 70$ meV of process I rather closely matches the optical gap of 550 cm$^{-1}$ ($\approx 68$ meV) found for $\delta$-Br by infrared spectroscopy [37]. As mentioned in section I, in Ref. [37] this gap was discussed in terms of excitations expected for a one-dimensional Wigner crystal. It was pointed out already long ago by Hubbard [38] that the periodic arrangement of electronic charges, caused by strong electronic repulsion in some quasi-one-dimensional organic salts, can be regarded as a "one-dimensional generalization of the classical Wigner lattice". Optically induced excitations of this lattice, essentially corresponding to the generation of domain-wall like defects in the charge-ordered chains, were already considered by Hubbard [38] and treated in more detail in Refs. [39,40]. It seems reasonable that the same excitations that are induced by infrared radiation in the optical experiments [37], can also arise via thermal activation and may then become detectable by dielectric spectroscopy. To rationalize this, it should be noted that these domain walls are generated by simply exchanging an electron between a neutral and positively charged molecule [37,39]. However, after being generated, these walls may also move, which involves just the same type of electron exchange as their generation [39]. As this electron transfer can be regarded as equivalent to the reorientation of a dipolar moment, it should be dielectrically active and may show up as relaxational process in dielectric spectroscopy, at least if it is slow enough to lie within the accessible frequency range. Obviously, for this Wigner lattice both the generation and the motion of a domain wall involve the same type of electron exchange and can be expected to be governed by the same energy barrier just as experimentally observed. Overall, it seems reasonable to assign the detected dielectric relaxation process I as documented in Figs. 1 and 4 to the thermally activated motion of the same excitations of the Wigner lattice as detected by optical spectroscopy in Ref. [37].

At first glance, relaxation times reaching around 0.1 s at the lowest temperatures as observed for process I (Fig. 4) seem surprisingly slow for electronic dynamics. However, one should be aware, that neutral and charged EDT-TTF-CONMe$_2$ molecules are known to exhibit different molecular conformations [28]. Therefore, these electron transfers can be assumed to be accompanied by conformational rearrangements of rather large molecules which should slow them down considerably. This non-canonical combined electronic and conformational type of motion may also be responsible for the mentioned unusually large pre-exponential factor of $\tau_0 \approx 10^{-8}$ s. Another interesting feature is the change of slope at high temperatures, observed for the log[$\tau(1/T)$] curve of relaxation process I (Fig. 4). Interestingly, it occurs rather close to the phase-transition temperature of 190 K. Thus it seems that this structural transition strongly affects the energy barrier for the electron transfer within the Wigner lattice.

It is an interesting fact that, when choosing a reasonable $\tau_0 = 3\times10^{-13}$ s, the extrapolated VFT curve used to fit the relaxation times of the glasslike ethylene-group dynamics (solid line in Fig. 4) also matches the three highest temperature points of process I detected by dielectric spectroscopy. Of course, one should be aware that such an extrapolation over about 10 decades is of limited significance. However, even for an extrapolation of the alternative Arrhenius law, indicated by the dash-dotted line in Fig. 4, a merging of both relaxations at high temperatures seems likely. A connection of both processes seems reasonable as both involve conformational dynamics: The glasslike relaxation process corresponds to conformational changes of significant parts of the organic molecules and the detected electronic domain-wall dynamics is accompanied by a conformational change of the molecules as a whole. Judging from Fig. 4, both processes seem to have comparable, rather fast relaxation times around $10^{-7}$ - $10^{-6}$ s at the highest temperatures $T > 200$ K and, only at lower temperatures, both dynamics become separated. Then the ethylene-group relaxation, occurring for all molecules, exhibits the typical cooperative glassy freezing marked by super-Arrhenius temperature dependence of the relaxation time [44,57,58,59,60]. In contrast, the defect-related domain-wall process, occurring for only a few molecules, is governed by conventional thermally activated behavior reflecting essentially independent relaxations of isolated dipolar degrees of freedom.

We want to mention that the behavior of the glasslike relaxation and process I, documented in Fig. 4, is reminiscent of the common $\alpha$-$\beta$-relaxation scenario often found for canonical glass formers [61,62], which would imply a completely different interpretation of process I: While the process revealed by thermal expansion corresponds to an $\alpha$ relaxation, governing the glassy freezing, within this scenario process I would be a so-called $\beta$ relaxation. As first shown by Johari and Goldstein [63], such $\beta$ processes seem to be an inherent property of the glassy state of matter. Various explanations of these Johari-Goldstein (JG) relaxations were proposed, e.g., in terms of "islands of mobility" [63,64] or small-angle reorientations [65,66]. The relaxation times of JG relaxations are known to show weaker temperature dependence than the $\alpha$ relaxation and to follow thermally activated behavior in extended temperature ranges just as it is the case for process I (Fig. 4). Finally, JG relaxations tend to merge with the $\alpha$ relaxation at high temperatures, i.e., their relaxation times approach each other under heating and, at high frequencies, only a single relaxation process is detected [61,62], which also is consistent with the present results.

Irrespective of the scenario being considered, one may ask whether the glasslike relaxation is also revealed by the dielectric experiments. Being the slowest process in the system, in principle it should lead to separate peaks in the $\sigma'(T)$ curves shown in Fig. 1(b) that are located at higher temperatures than the two clearly visible peaks in this figure. Assuming that $\tau(T)$ indeed follows the VFT law shown in Fig. 4 and using the condition $\tau(T_p) = 1/(2\pi\nu)$, the arrows in Fig. 1(b) indicate the expected $\alpha$-peak positions for selected frequencies. While no clear additional peaks can be discerned at these temperatures, the found gradual decrease towards high temperatures in this region may well be consistent with the presence of a broad additional slow process, superimposed by process I. (Unfortunately, in the relevant temperature range of about 150 - 250 K, reliable loss data are only available for $\nu \geq 13.1$ kHz as the dielectric measurements were hampered by the non-ideal needle-like shape of the crystals.)

Currently, we only can speculate about the microscopic origin of relaxation II, seen at temperatures below 100 K and representing the fastest process detected in the present work. For the specific system treated in Ref. [39], it was shown that second-neighbor hopping may cause additional absorption features in the optical spectra, also involving excitonic states arising from the Coulomb attraction between domain walls. Another explanation of this process could be additional dipolar-active intramolecular modes. Finally, there are various hydrogen bonds (intra- and interstack) in $\delta$-Br [41]. In some materials, relaxational response is found to arise from motions of protons in double-well potentials, arising at H-bonds [67,68]. As relaxation II has a clearly higher relaxation strength $\Delta\varepsilon$ than the defect-related process I [cf. Fig. 1(a)], the latter two explanations seem more likely.

## V. SUMMARY AND CONCLUSIONS

In summary, by combining thermal-expansion and dielectric investigations we have found a rich relaxational response of the charge-ordered Mott insulator $\delta$-Br. The thermal-expansion measurements provide clear evidence for relatively slow relaxational dynamics in $\delta$-Br which exhibits glassy freezing with a glass temperature of about 130 K. It most likely arises from the same configurational degrees of freedom of the organic molecules as previously identified for various related organic charge-transfer salts [1,29,31,34,47]. It is of special significance that here we have observed such a glasslike behavior in a quasi-one-dimensional system with spherical anions. In contrast, in the previously investigated compounds the anions form polymerlike chains and reorientational motions of these chains were suggested to be involved in the glasslike transition in these systems [33], which clearly can be excluded for $\delta$-Br.

Moreover, by dielectric spectroscopy we find evidence for two additional, much faster relaxation processes at temperatures between about 20 and 250 K. For the origin of the slower relaxation (process I), we propose the same theoretically predicted excitations of one-dimensional Wigner lattices as previously found in this material by infrared spectroscopy [37]. The relaxation times of the glasslike relaxation and process I seem to merge at high temperatures, probably reflecting the involvement of conformational changes of the organic molecules in both cases. However, below about 200 K the typical cooperative nature of the glasslike process causes a strong super-Arrhenius temperature dependence of its relaxation time, while the defect-related process I reveals a much more gradual slowing down due to its single-dipole nature and obviously also due to the occurrence of the structural transition at 190 K. It should be clearly noted that this scenario is tentative and additional measurements with different methods, closing the dynamic gap between the dielectric and thermal-expansion investigations, may be helpful to prove its validity.

Compared to the BEDT-TTF compounds previously investigated by us [11,20,21], $\delta$-Br exhibits no ferroelectric transition in the investigated temperature range, while the non-canonical increase of the dielectric constant at high temperatures [Fig. 1(a)] may indicate such a transition at higher temperatures. The lack of dimerization and the rather weak interstack coupling makes this material a rather good realization of a one-dimensional Wigner-lattice and leads to the corresponding relaxational dynamics which is absent in the two-dimensional BEDT-TTF systems.

## ACKNOWLEDGMENTS

This work was supported by the Deutsche Forschungsgemeinschaft through the Transregional Collaborative Research Centers TRR 49 and TRR 80.

*Corresponding author: Jonas.Fischer@Physik.Uni-Augsburg.de
#Both authors contributed equally to the present paper.